\documentclass[letter]{aa}
\usepackage{pst-all}
\usepackage{color}
\usepackage{txfonts}
\usepackage{makecell}
\usepackage{amsmath}
\usepackage{amssymb}
\usepackage{bm}
\usepackage{graphicx}
\usepackage{array,multirow}
\usepackage{url}
\usepackage{layouts}
\usepackage{siunitx}
\usepackage[pdfauthor={Barbosa et al.},
            pdftitle={What does not drive the IMF of ETGs}]{hyperref}
\hypersetup{colorlinks=true, linkcolor=blue, citecolor=blue, urlcolor=blue}
\usepackage[flushleft]{threeparttable}
\usepackage{acronym}
\usepackage{soul}

\DeclareSIUnit\ergon{erg}
\bibpunct{(}{)}{;}{a}{}{,} 

\acrodef{ETGs}{early-type galaxies}
\acrodef{IMF}{initial mass function}
\acrodef{SFH}{star formation history}
\acrodef{SNR}{signal-to-noise ratio}
\acrodef{SSP}{single stellar population}
\acrodef{SED}{spectral energy distribution}

\definecolor{darkgreen}{rgb}{0.09, 0.45, 0.27}

\defcitealias{2016A&A...589A.139B}{B16}
\defcitealias{2018A&A...609A..78B}{B18}
\begin{document} 

\title{What does (not) drive the variation of the low-mass end \\ of the stellar initial mass function of early-type galaxies}
\subtitle{\textit{MUSE Paintbox II}}
\titlerunning{What does (not) drive variation in initial mass function of ETGs}
\author{
C.~E. Barbosa\inst{\ref{usp}}\fnmsep\thanks{Corresponding author: {\tt carlos.barbosa@usp.br}.} \and
C. Spiniello\inst{\ref{oxf},\ref{inaf_naples}} \and
M. Arnaboldi\inst{\ref{eso}} \and
L. Coccato\inst{\ref{eso}} \and
M. Hilker\inst{\ref{eso}} \and
T. Richtler\inst{\ref{concepcion}}
              }

\institute{
Universidade de S\~ao Paulo, IAG, Departamento de Astronomia, Rua do Mat\~ao 1226, S\~ao Paulo, SP, Brazil\label{usp}
\and
Department of Physics, University of Oxford, Denys Wilkinson Building, Keble Road, Oxford OX1 3RH, UK\label{oxf} 
\and 
INAF, Osservatorio Astronomico di Capodimonte, Via Moiariello  16, 80131, Naples, Italy\label{inaf_naples}
\and
European Southern Observatory,  Karl-Schwarzschild-Stra\ss{}e 2, 85748, Garching, Germany\label{eso}
\and
Departamento de Astronomia, Universidad de Concepci\'on, Concepci\'on, Chile\label{concepcion}     
}

   \date{Received October 30, 2020; Accepted December 1, 2020}

  \abstract
   {The stellar \ac{IMF} seems to be variable and not universal, as argued in the literature in the last three decades. Several relations among the low-mass end of the IMF slope and other stellar population, photometrical or kinematical parameters of massive \ac{ETGs} have been proposed, but a consolidated agreement on a factual cause of the observed variations has not been reached yet.} 
   {We investigate the relations between the IMF and other stellar population parameters in NGC~3311, the central galaxy of the Hydra I cluster. NGC~3311 is a unique laboratory, characterized by old and metal-rich stars, like other massive ETGs for which the IMF slope has been measured to be bottom-heavy (i.e. dwarf-rich), but has unusual stellar velocity dispersion and [$\alpha/$Fe] profiles, both increasing with radius.} 
   {We use the spatially resolved stellar population parameters (age, total metallicity, [$\alpha$/Fe]) derived in a companion paper (Barbosa et al. 2020) from Bayesian full-spectrum fitting of high signal-to-noise MUSE observations to compare the IMF slope in the central part of NGC~3311 ($R \lesssim 16$ kpc) against other stellar parameters, with the goal of assessing their relations/dependencies.}
   {For NGC~3311, we unambiguously invalidate the previously observed direct correlation between the IMF slope and the local stellar velocity dispersion, confirming some doubts already raised in the literature. This relation may arise as a spatial coincidence only, between the region with the largest stellar velocity dispersion value, with that where the oldest, \textit{in situ} population is found and dominates the light. We also show robust evidence that the proposed IMF-metallicity relation is contaminated by the degeneracy between these two parameters. We do confirm that the stellar content in the innermost region of NGC~3311 follows a bottom-heavy IMF, in line with other literature results. The tightest correlations we found are those between stellar age and IMF and between galactocentric radius and IMF. }
   {The variation of the IMF at its low-mass end is not due to kinematical, dynamical, or global properties in NGC~3311. We speculate instead that the IMF might be dwarf-dominated in the "red-nuggets" formed through a very short and intense star formation episode at high redshifts ($z>2$), when the universe was denser and richer in gas, and that then ended up being the central cores of today giant ellipticals.}


   \maketitle
%

\section{Introduction}
The \ac{IMF} is the distribution of stars per unit of mass formed in a volume of space within a single star-formation event. Since the mass of a star determines its subsequent evolutionary path, the \ac{IMF} influences all the observable properties of a stellar system. 
In addition, the \ac{IMF} slope at the low-mass end modifies the stellar mass-to-light ratio value in any galaxy. In fact, low-mass stars account for more than half of the mass budget while they contribute very little to the integrated luminosity of a galaxy dominated by an old stellar population \citep[see, e.g., Fig.~2 in ][]{2012ApJ...747...69C}.  
This makes the characterization of the low-mass \ac{IMF} slope from integrated light a challenging task, but crucial to assess the relative contribution of stars and dark matter,  and to understand their relative distribution in the baryon-dominated inner galaxy regions. 

In the last decade, a large number of studies reported that the low-mass end of the \ac{IMF} slope in massive \ac{ETGs} is bottom-heavier (i.e. dwarf-richer) in relation to the IMF inferred in the Milky Way (MW), as indicated by a variety of methods, such as stellar populations, gravitational lensing and dynamics \citep[see][]{2003MNRAS.339L..12C, 2010Natur.468..940V, Treu+10, 2012ApJ...760...71C, 2012ApJ...753L..32S,2014MNRAS.438.1483S, 2012Natur.484..485C,  2013MNRAS.433.3017L, 2017MNRAS.464.3597L}. 
However, a general consensus on which physical quantity is really responsible for the non-universality of the IMF is far from being reached. The stellar velocity dispersion \citep{2012Natur.484..485C, 2014MNRAS.438.1483S}, the total mass density \citep{Spiniello+15}, the metallicity \citep{2015ApJ...806L..31M}, the magnesium abundance \citep{2017MNRAS.464.3597L}, the age and the $\alpha$-element abundance \citep{2014ApJ...792L..37M} have been investigated so far. 

More recently, spatially resolved studies have found that the bottom-heavy \ac{IMF} is concentrated in the innermost region of galaxies \citep{2015MNRAS.447.1033M, 2017ApJ...841...68V, 2018MNRAS.478.4084S, 2018MNRAS.477.3954P,2018MNRAS.479.2443V, 2019MNRAS.489.4090L}, indicating that also local properties, not only global, may drive variation in the low-mass end of the \ac{IMF}. These findings are particularly interesting in the context of the two-phase formation scenario \citep{2009ApJ...699L.178N, 2010ApJ...725.2312O, 2012MNRAS.425.3119H, 2016MNRAS.458.2371R, 2020arXiv200901823P}, as the center is where the \textit{in situ} population, formed during the first phase of the mass assembly dominates the light.

The majority of massive \ac{ETGs} shows mostly flat or falling stellar velocity dispersion ($\sigma_*$) profiles \citep{1995ApJ...438..539F, 1996AJ....112..797S, 1999MNRAS.307..131C, 2007MNRAS.378.1507B, 2014MNRAS.445L..79E}.   
Consequently, the IMF has always been measured to be bottom-heavy in the region where the $\sigma_*$ profile had the central peak. The same holds for the $[\alpha$/Fe] abundance, which is often larger in the center than in the outer regions. Thus it has not been possible until now to assess independently whether the spatial variation of the low-mass end of the IMF is driven by the kinematics (i.e. the IMF is dwarf-rich wherever $\sigma_{\star}$ is high), the stellar populations (i.e. the IMF is dwarf-rich when the metallicity/[$\alpha$/Fe] is super solar) 
or rather by other mechanisms linked to the intense, early star formation (i.e. the IMF is dwarf-rich in the innermost region, where the \textit{in situ} stars dominate, despite of the local kinematics). 

In this context, NGC~3311, the central galaxy of the Hydra~I cluster, is a perfect ``laboratory'' to disentangle between these scenarios, having both a $\sigma_{\star}$ and a $[\alpha$/Fe] profile rising with increasing galactocentric distances\footnote{We note that few other massive galaxies with rising $\sigma_*$ profiles exist, see e.g. \citet{2017MNRAS.464..356V}.}, while the total metallicity is instead higher in the center \citep[][hereafter B16]{2016A&A...589A.139B}. The very peculiar profiles in NGC~3311 are the results of the presence of multiple components with distinct kinematics and stellar populations, as we show in detail in the companion paper MUSE Paintbox I (Barbosa et al.~2020, hereafter B20). There we performed a fully Bayesian, full-spectral fitting analysis on high signal-to-noise MUSE spectra and obtained accurate spatially resolved measurements of light-weighted age, metallicity ([Z/H]), $\alpha$-element abundance ($[\alpha$/Fe]) and IMF slope ($\Gamma_b$). 
We refer the reader to B20 for more details about the stellar population modeling. 
Here we focus mainly on the low mass-end of the IMF slope and its possible correlations with the other kinematical and stellar population parameters. We note that the spatially resolved stellar population properties of NGC~3311 cover a range of metallicities, [$\alpha$/Fe] abundances and $\sigma_{\star}$ that would allow us to observe IMF gradients and correlations, assuming the IMF depends on these parameters, in line with previous studies.

\section{What drives the IMF variations?}
\label{sec:imf}

In B20, we use the E-MILES \citep{2016MNRAS.463.3409V} stellar population (SP) models with a ``bimodal'' IMF \citep{1996ApJS..106..307V}, which is obtained by changing the high-mass end ($M \gtrsim 0.6M_{\odot}$) power-law slope ($\Gamma_b$). 
The \ac{IMF} is then normalized to $1M_\odot$, as more massive stars are already dead in old stellar populations, hence a change in the $\Gamma_b$ slope can be interpreted as a change in the ratio between stars with masses above and below $0.6M_\odot$. 
In practice, we are constraining the "present-day" mass function between $0.1$ and $1$ solar masses.

In the next sections, we try to understand what are the possible physical causes for the non-universality of the \ac{IMF} in NGC~3311, as well as in other massive ETGs. 
To this purpose, we compare the results from NGC~3311 (obtained in B20, 
shown in all figures as data-points color-coded by distance from the center) with those obtained globally across different galaxies (dashed lines in all plots) and spatially within one or more objects (solid lines in all plots). 

\begin{figure}[!t]
\centering
\includegraphics[width=\columnwidth]{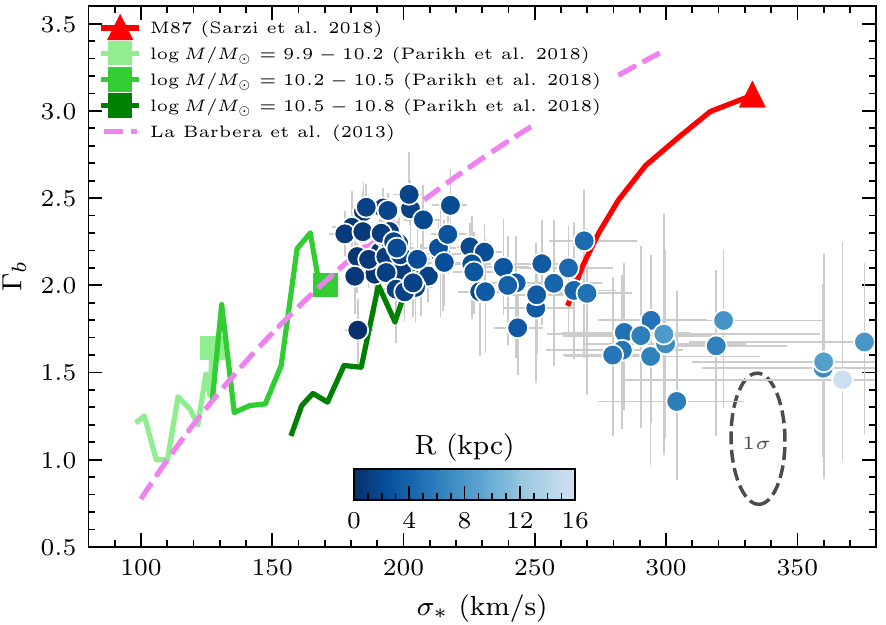}
\includegraphics[width=\columnwidth]{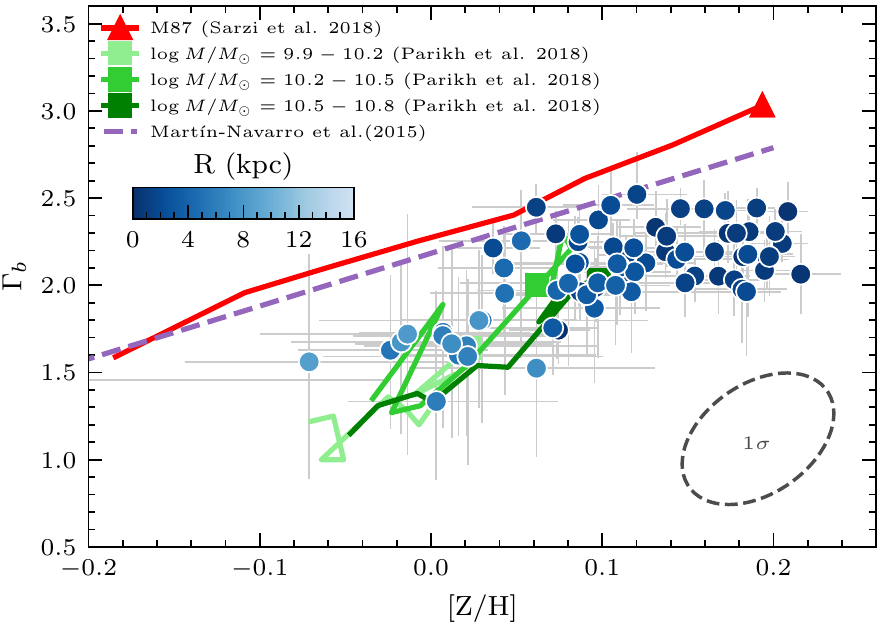}
\caption{IMF slope ($\Gamma_b$) versus stellar velocity dispersion ($\sigma_{\star}$, \textit{upper panel}), and total metallicity ([Z/H], \textit{bottom panel}). Blue circles indicate data-points for NGC~3311, colored according to the galactocentric distance. For other works, solid lines refer to local variations within galaxies (with a symbol indicating the center of the objects), and dashed lines refer to global variation across galaxies.}
\label{fig:imf_sigma_Z}
\end{figure}

\subsection{Correlations that do not hold for NGC~3311}
\label{sec:not_drivers}
The stellar velocity dispersion has been one of the first ``drivers'' proposed to explain the bottom-heaviness of the IMF both across different galaxies \citep{2010Natur.468..940V, Treu+10, 2012ApJ...760...71C,2012Natur.484..485C, 2014MNRAS.438.1483S} and within individual objects \citep{2015MNRAS.447.1033M, 2018MNRAS.478.4084S, 2019MNRAS.489.4090L}. However, recent indications have aroused questioning the IMF-$\sigma_{\star}$ relation, mainly based on the fact that the spatial gradients of the two quantities were often very different in radial extension (e.g. \citealt{2018MNRAS.478.4084S}).
But, above all, the IMF slope has never been measured in galaxies with rising $\sigma_*$ and [$\alpha$/Fe] profiles like NGC~3311. 

In the upper panel of  Figure~\ref{fig:imf_sigma_Z}, we decidedly show that a direct linear correlation between the IMF bottom-heaviness and $\sigma_{\star}$ 
does not hold for NGC~3311. There, we compare the results of B20 with those obtained for M87 \citep[red line, ][]{2018MNRAS.478.4084S}, and from a sample of MaNGA \citep{2015ApJ...798....7B} galaxies, stacked in three mass bins  \citep[green lines, ][]{2018MNRAS.477.3954P}.  We also show the global results obtained in \citet{2013MNRAS.433.3017L} based on a spectroscopic study on a large sample ($\sim24000$) of ETGs from the SPIDER Survey \citep{2009arXiv0912.4547L}, performed with the same stellar population models (and the same IMF) used in B20. 
While a clear direct correlation is visible for all previous studies, a mild anti-correlation is inferred for NGC~3311, with the IMF slope being more bottom-heavy for points with lower $\sigma_{\star}$. However, the IMF is dwarf-rich in the center of NGC~3311. As we will further show in the next section, all the results support a conjecture where the innermost region of massive galaxies is characterized by a dwarf-rich IMF, despite their kinematical properties. 

Another parameter that has been reported to correlate directly with the IMF slope is the total metallicity ([Z/H],  \citealt{2015ApJ...806L..31M}). 
Tentatively, a correlation can be seen for NGC~3311 in the lower panel of Figure~\ref{fig:imf_sigma_Z}, which is in agreement with the global results presented by the CALIFA Collaboration in \citet{2015ApJ...806L..31M}, and with the spatially resolved results obtained for M87 and the MaNGA stacked galaxies. These observational results suggest that the IMF is bottom-heavy for a more metal-rich population.  
Although we reproduce this relation, we also show that the slope of the correlation is almost perfectly aligned with the mean posterior distribution between the two parameters obtained in B20, shown as an ellipsoidal dashed line in the Figure. Therefore, there is strong ground to believe that the relation is at least partially due to an observational degeneracy between the two plotted quantities. \footnote{This evidence reinforces the importance of committing to a fully Bayesian approach and carefully inspecting the posterior probability results, not simply as a way to determine point estimates and uncertainties, but also as a tool to understand our limitations from an observational perspective.}

Furthermore, from a theoretical perspective, simulations have demonstrated that the trends predicted by the IMF-[Z/H] correlation fail to reproduce the general trend observed between the iron and $\alpha$-element abundance \citep{2019MNRAS.482..118G}. Finally, we also note that other recent observational results \citep[e.g.,][]{2019A&A...626A.124M} indicate that the metallicity does not trace the \ac{IMF} in spatially resolved observations\footnote{The results in \citet{2019A&A...626A.124M} suggest that, at least for the galaxy FCC 167, the metallicity track the distribution of cold orbits whereas the IMF slope that of warm orbits.}. 

\subsection{Correlations that do hold for NGC~3311}
\label{sec:drivers}
\begin{figure}[!h]
\centering
\includegraphics[width=\columnwidth]{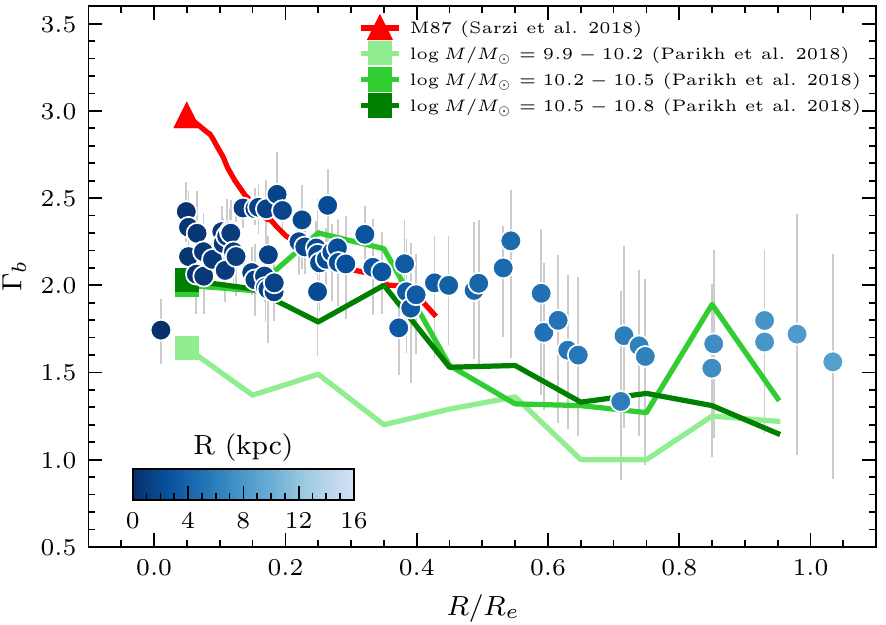}
\includegraphics[width=\columnwidth]{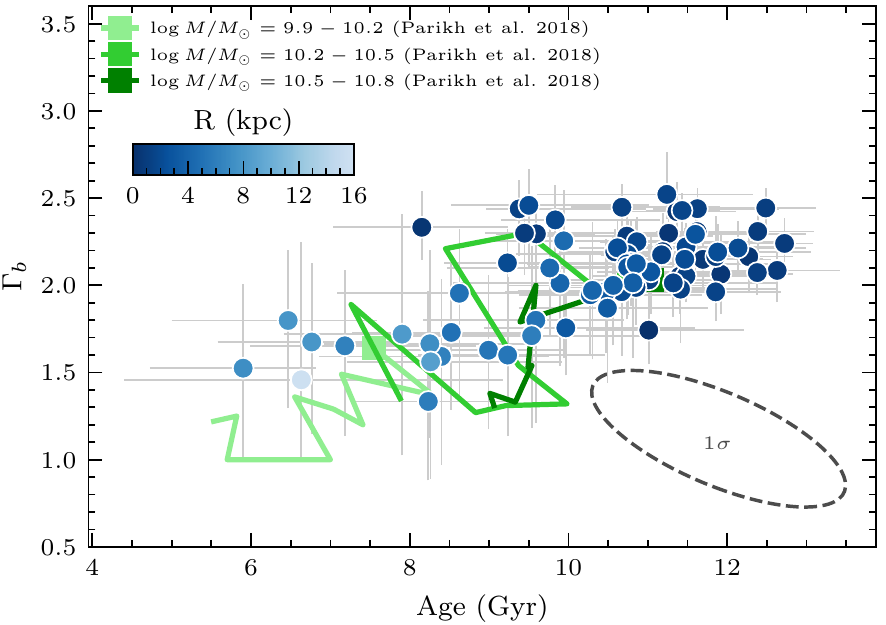}
\includegraphics[width=\columnwidth]{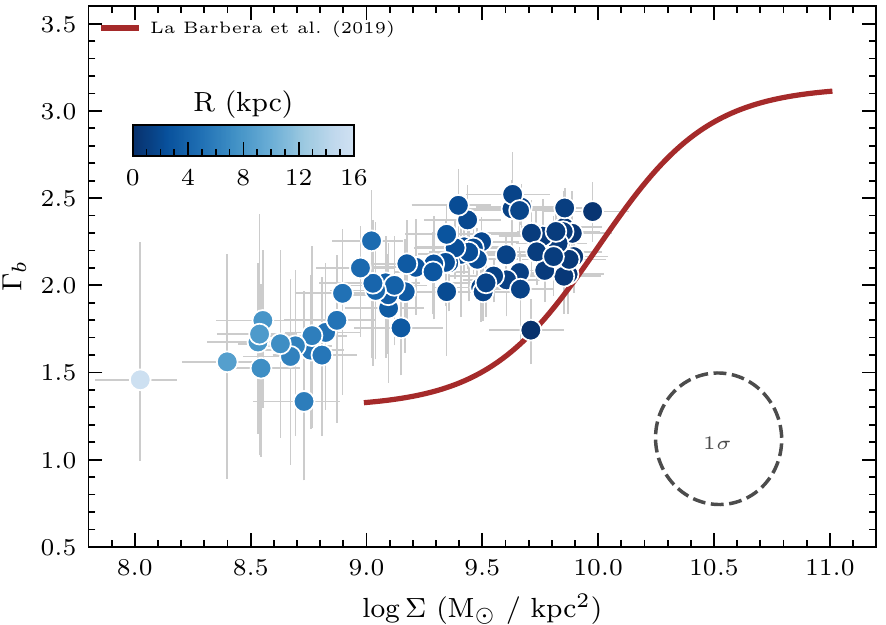}
\caption{The IMF spatial gradient (\textit{upper panel}), and the correlation between $\Gamma_b$ and the stellar age (\textit{middle panel}) and between $\Gamma_b$ and the total stellar density (\textit{bottom panel}), all pointing towards the result that the IMF slope is bottom-heavy only for the "in situ" stars, formed at very high redshift from the first-phase of the two phase formation scenario. }
\label{fig:relations}
\end{figure}

In Figure~\ref{fig:relations} we plot the observed relations between the IMF slope and radius (normalized by $R_e$, top panel), stellar age (middle panel) and surface stellar density (bottom panel).
As in the previous figures, we compare the data points showing the results on NGC~3311 (obtained in B20) with lines showing the relations reported in the literature. 
Interestingly, these three plots support the following evidence: despite the different kinematical properties of the galaxies' central regions, the IMF is always dwarf-rich within the innermost few kilo-parsecs. There, according to the so-called two-phase formation scenario, the old and very dense ``red-nugget'' formed during the first phase of the mass assembly dominates the light. 

In line with what has been suggested by other authors \citep{2017ApJ...841...68V, 2019A&A...626A.124M, 2020ARA&A..58..577S}, we, therefore, argue that the excess of dwarf stars may have originated from the first phase of galaxy formation
in the early Universe, characterized by high density, temperature, and turbulence of the gas. These are, according to theoretical works \citep{2012MNRAS.423.2037H, 2014ApJ...796...75C}, key parameters driving the fragmentation of molecular clouds (higher density and temperature should make the fragmentation easier, forming more dwarf stars, i.e. a bottom-heavier IMF)\footnote{Very recently a further parameter has been advocated by \citet[]{2020MNRAS.498.4051D}: the abundance of deuterium in the birth clouds of forming stars}. 
Later accretions of less massive satellites with an “MW-like” IMF (because formed later on, under less extreme temperature and density conditions), would settle preferentially in the galaxy periphery due to the lower densities and masses of the accreted systems and thus create spatial IMF gradient \citep{2017A&A...603A..38S, 2020arXiv200901823P}\footnote{In NGC~3311, the later mass accretion scenario is supported by the rising velocity dispersion profile that has been considered as the ``smoking gun''  of recent accretion of low-mass systems, dominating the dispersion scatter and leading to larger $\sigma_{\star}$ at larger distances \citep{2018A&A...619A..70H}.}. 
Finally, our findings are consistent with those reported in \citet{2012ApJ...760...71C} and \citet{2019MNRAS.489.4090L}, 
who found that the IMF becomes increasingly bottom-heavy when either the star formation timescale becomes shorter (see \S\ref{sec:alpha}),  or the surface density and/or the interstellar medium pressure increase.

\subsection{The (apparently) puzzling case of the [$\alpha$/Fe] abundance}
\label{sec:alpha}

According to the scenario described above, and in line with the results of B20, the central core of NGC~3311 is a "red-nugget" that formed its stars under extreme conditions. 
Such a short star formation time-scale would lead to a high [$\alpha$/Fe] abundance \citep{2005ApJ...621..673T}. 
This is explained by the fact that the star formation quenching took place before the Type Ia supernova explosions were able to pollute the interstellar medium with iron.

\begin{figure}
\centering
\includegraphics[width=\columnwidth]{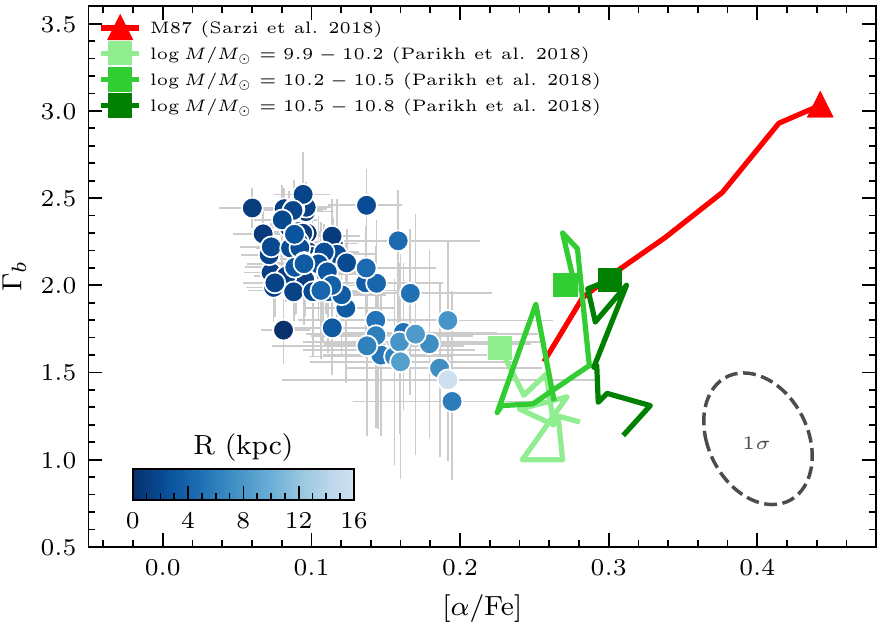}
\caption{IMF slope - [$\alpha$/Fe] relation. For NGC~3311 (blue points) these two quantities are anti-correlated. This is the opposite of what happens for M87 (red). However, we stress that in B20 we used full-spectral fitting, rather than line-indices as in \citet{2018MNRAS.478.4084S} for M87, thus measuring a different quantity (from indices one is much more sensitive to Mg than to the "bulk" changes in all $\alpha$ elements). Note that the [$\alpha$/Fe] estimates obtained in \citet{2018A&A...609A..78B} from the same data but using line-indices imply a high value, which would shift the blue data-points towards the red and green lines. The situation is much more complicated for the results based on MaNGA (green lines) where different galaxies, with possibly different  $[\alpha$/Fe] profiles are stacked together. }
\label{fig:alphafe}
\end{figure}

The $\alpha$-abundance of stars in the innermost region of NGC~3311 is not extremely high ($0.08\pm0.02$, B20). In principle, this might question our conclusions that the stars in the core of NGC~3311 have been formed via a burst during the first phase of the two-phase scenario and with a dwarf-rich IMF. However, as showed in Fig~8 of B20, the NCG~3311 core requires a very large [Mg/Fe] ($>0.3$), which is, in practice,  what was measured in \citet{2005ApJ...621..673T}, since they derived [$\alpha$/Fe] from the absorption line indices Mg$b$ and $\langle$Fe$\rangle$. 
The same is valid for the measurements performed in \citet{ 2014ApJ...792L..37M}, which also proposed a direct global correlation between the IMF slope and the [$\alpha$/Fe] but actually measured [Mg/Fe] instead\footnote{We refer the reader to B20 for more details about the different weight one gives to the $\alpha$-elements when using line-index or when using full spectral fitting.}.  
Hence, once again, the direct correlation found between the IMF slope and the [$\alpha$/Fe] (or, more exactly, [Mg/Fe]) in other ETGs, e.g. M87, is found because for the majority of the massive ETGs the region where the [Mg/Fe] is maximum spatially coincide with the region where the \textit{in-situ} pristine population of stars dominates the light. 
We can now prove such a statement thanks to the unusual mass assembly of NGC~3311 (see B20 for more details), which caused an inverse $\alpha$-element profile. In fact, in our case, $\Gamma_b$ decreases when [$\alpha$/Fe] increases, as shown in  Figure~\ref{fig:alphafe}. We stress that the offset between our results and those of \citet{2018MNRAS.478.4084S} and \citet{2018MNRAS.477.3954P} is due to the fact that we measure variations in all the $\alpha$-elements, rather than only Mg. The [Mg/Fe] in the innermost region of NGC~3311 is indeed large (see \citealt{2018A&A...609A..78B} and B20), thus these stars are compatible with having been formed via a burst of star formation. Finally, we note that the large $[\alpha$/Fe] measured in the galactic halo of NGC~3311 is not caused by the fact that these stars have been formed at high-z "in situ" via a burst, but rather by the late accretion of previously quenched satellites (see B20 for more details).

In conclusion, all evidence points consistently to the fact that the IMF is dwarf-rich when stars are formed through a high-z intense and brief star formation episode. 

\section{Summary and Conclusion}
\label{sec:conclusion}
In this letter, we used spatially resolved kinematical and stellar populations results from NGC~3311, the central galaxy of the Hydra~I cluster to address the question of whether the relations between the low-mass end of the IMF slope and kinematical ($\sigma_{\star}$) or stellar population (age, [Z/H], [$\alpha/$Fe]) parameters are real, physically motivated relations. 
We find that the IMF-$\sigma_{\star}$ relation is mainly a spatial coincidence arising from the fact that the majority of massive \ac{ETGs} show a $\sigma_{\star}$ profile that is peaked in the center likely because of violent relaxation \citep{2012MNRAS.425.3119H}. This is not true for NGC~3311, but its core IMF is nonetheless bottom heavy. 
The strength of the IMF-metallicity direct correlation may instead be augmented by the degeneracy between $\Gamma_b$ and [Z/H], as addressed more extensively in the companion paper Barbosa et al. 2020. 

We also presented our arguments in support of a two-phase formation scenario with two IMF with different low-mass end slope for the two distinct phases, as suggested by \citet{2020ARA&A..58..577S}. This scenario is able to reconcile all the different observational and theoretical evidence in a spatially resolved sense, within a single galaxy, as well as in a "global" sense on large galaxy samples. 
Studying the IMF spatial gradients and the relation linking its slope $\Gamma_b$ to the stellar age and that between $\Gamma_b$ and stellar surface density, we explain the bottom-heaviness of the IMF as characteristics of the "relic" nucleus which was formed during the first phase, at very high-z in a violent, short, star formation episode ($\tau\sim 100 {\rm Myr , \,\, SFR} \ge 10^{3} {\rm M}_{\odot} {\rm yr}^{-1}$).  
The accreted stellar component,  which dominates the light budget at larger galactocentric radii and was formed under less extreme and more time-extended star formation episodes (second phase), is instead described by a standard, ``MW-like'' IMF. 

This version of the two-phase formation scenario naturally explains the radial IMF gradients recently reported for ``normal'' massive ETGs, but the very recent results obtained on ``relic galaxies'' provide further, independent support to this idea. Relics are rare local ultra-compact massive red and dead galaxies, which missed the second phase of accretion, thus evolving passively and unperturbed after the first burst of star formation. Since they missed this second phase, relics would have a bottom-heavy IMF everywhere, which is exactly what has been reported by \citet{2015MNRAS.451.1081M} and \citet{2017MNRAS.467.1929F}  on the only three relics systems for which spatially resolved stellar population analysis has been performed so far. The outlook is bright, thanks to the newly started \mbox{{\tt INSPIRE}} Project \citep{2020arXiv201105347S} which aims at confirming a larger number of relics at $z<0.5$ and precisely infer their IMF slope thanks to high signal-to-noise ratio X-Shooter spectra.

\vspace{1cm}
\begin{acknowledgements}
The authors wish to acknowledge the anonymous referee for the constructive report.  The authors acknowledge C. Mendes de Oliveira for discussions and suggestions. CEB gratefully acknowledges the S\~ao Paulo Research Foundation (FAPESP), grants 2011/51680-6,  2016/12331-0, 2018/24389-8.  CS is supported by a Hintze Fellowship at the Oxford Centre for Astrophysical Surveys, which is funded through generous support from the Hintze Family Charitable Foundation. TR acknowledges support from the BASAL Centro de Astrofisica y Tecnologias Afines (CATA) PFB-06/2007. \\
This work is based on observations collected at the European Organisation for Astronomical Research in the Southern Hemisphere under ESO programme 094.B-0711(A). 
It has made use of the computing facilities of the Laboratory of Astroinformatics (Instituto de Astronomia, Geof\'isica e Ci\^encias Atmosf\'ericas, Departamento de Astronomia/USP, NAT/Unicsul), whose purchase was made possible by FAPESP (grant 2009/54006-4) and the INCT-A. This research has made use of the NASA/IPAC Extragalactic Database (NED), which is operated by the Jet Propulsion Laboratory, California Institute of Technology, under contract with the National Aeronautics and Space Administration.
\end{acknowledgements}

\bibliographystyle{aa.bst} 
\bibliography{biblio.bib}

\end{document}